\documentclass[superscriptaddress,twocolumn]{revtex4}
\usepackage[tbtags]{amsmath}
\usepackage{amsfonts}
\usepackage{mathrsfs}
\usepackage{amssymb}
\usepackage{graphicx}
\usepackage{epsfig,graphicx,times}
\usepackage{subfigure}
\usepackage{color}
\usepackage{CJK}

\begin{document}

\title{Routing surface plasmons by a quantum-dot nanostructure: nonlinear dispersion effects}

\author{Jin-Song Huang\footnote{jshuangjs@126.com} }

\affiliation{ School of Information Engineering, Jiangxi University
of Science and Technology, Ganzhou 341000, China}

\author{Jia-Hao Zhang}

\affiliation{ School of Information Engineering, Jiangxi University
of Science and Technology, Ganzhou 341000, China}

\author{L. F. Wei\footnote{weilianfu@gmail.com}}
\affiliation{Information Quantum Technology Laboratory, School of Information Science
and Technology, Southwest Jiaotong University, Chengdu 610031,
China}
\affiliation{State Key Laboratory of Optoelectronic Materials
and Technologies, School of Physics, Sun Yat-Sen
University Guangzhou 510275, China}

\begin{abstract}

 Usually, the liner waveguides with single quantum emitters are utilized
as routers to construct the quantum network in quantum information
processings. Here, we investigate the influence of the nonlinear
dispersion on quantum routing of single surface plasmons, between
two metal nanowires with a pair of quantum dots. By using a full
quantum theory in real space, we obtain the routing probabilities of
a single surface plasmon into the four outports of two plasmonic
waveguides scattered by a pair of quantum dots. It is shown that, by
properly designing the inter-dot distance and the dot-plasmon
couplings, the routing capability of the surface plasmons between
the plasmonic waveguide channels can be significantly higher than
the relevant network formed by the single-emitter waveguides with the linear
dispersions. Interestingly, the present quadratic dispersions in the
waveguides deliver the manifest Fano-like resonances of the
surface-plasmon transport. Therefore, the proposed double-dot
configuration could be utilized as a robust quantum router for
controlling the surface-plasmon routing in the plasmonic waveguides
and a plasmonic Fano-like resonance controller.


\end{abstract}
\maketitle

Quantum nodes are the key elements of the quantum
network~\cite{Kimble}, since they coherently connect different
quantum channels of the network. Quantum routers in the nodes,
provided usually by these systems of waveguides and quantum
emitters, are utilized to control the propagating paths of the
quantum signals, wherein the photons are served as the ideal
carriers of the information in these quantum channels, due to their
long-lived coherences and less dissipations~\cite{Brien}.

Recently, many
theoretical~\cite{Zhou,Lu,Yan,Agarwal,Xia,Lemr,wei2,wei3,Lu2}
and experimental~\cite{Aoki,Hoi,Shomroni} investigations have been devoted to
the quantum routings of single photons in various optical quantum
networks. Typically, Zhou et al and Lu et al demonstrated how to
realize the quantum routing of a single photon in the X-shaped
coupled-resonator waveguides by using the scattering of a
three-level atom. It showed that, with the help of a classical
driving~\cite{Zhou,Lu}, single photons incident from one channel
can be transferred into another one. Based on the photonic
interferences related to the different phases acquired in different
paths, Yan et al~\cite{Yan} examined that single-photon
routing can be realized in a network with multiple input and output
ports. However, low routing rate (no more than 0.5) of the single photons from the input channel into another channel exists in these single-emitter configurations, due to the emission asymmetry of the quantum emitter, which may restrict its more potential applications. For a multichannel quantum network, it is thus of considerable interest to design a quantum router with high routing capability.

Surface plasmons(SPs), known as the collective excitations of the
electrons on metal-dielectric interfaces, play an
important role in confining the electromagnetic wave at a nanometer
scale, as the important information carriers in various
functional nanophotonic devices and integrated optical
systems~\cite{Tame}. Additionally, SPs provide many potential
applications in quantum computation and quantum information
processing, since they reveal strong analogies to the photons
propagating along the conventional dielectric optical
components~\cite{Zia,Yu,Savel}. With the recent experimental
developments, an integrated system of a single metal nanowire
coupled with quantum dots (QDs) has been fabricated successfully as a
plasmonic waveguide~\cite{Akimov,Fedutik}, which provides an idea platform
to investigate the coherent transport of the SPs.
Consequently, wide attention has been addressed to the
transport properties of the SPs coupled to the waveguides in the linear dispersion regime,  and transistors~\cite{Chang,Hong},
switching~\cite{Kim1}, and scattering grating~\cite{Li} devices of SPs
have been designed accordingly. Within such a device, the
Fano resonance~\cite{Chen1,Chen2,Cheng1} and photonic
entanglements~\cite{Chen3,Chen4} have also been demonstrated.
However, the transport of the SPs with nonlinear dispersion relations has been paid relatively-less attention, although certain interesting behaviors, e.g., the four-peak
structure in the reflected spectrum~\cite{Cheng3}, jiggling
scattering behavior, and Fano-like resonance have been
predicted~\cite{Chen5}.

Motivated by the work mentioned above, we discuss how
to implement the high-efficient quantum routing of the SPs in a
nonlinear waveguide, which is made of two metal nanowires coupled to two separated QDs. The simple quantum network model, with two SPs channels and four ports, is
served as a quantum router of the SPs inputting from one of the
ports. By adopting a full quantum theory in real
space, the reflection and transmission
probabilities of single SPs into four ports are obtained exactly. It
has shown that, due to the quadratic dispersions in the double-dot
configuration, the routing capability of single SPs transferred from the incident
channel into another channel can reach to an extremely-high value,
by properly designing the inter-dot distance and the QD-SP
couplings. Furthermore, the jiggling behavior and multiple Fano-like
shapes of the scattering spectra can be observed. This implies
that the proposed system may be served as a robust quantum
router and a Fano-like resonance controller of the SPs for various
integrated quantum optical applications.

\section{Theoretical mode and its exact solution}

As shown in Fig.~1, the considered system consists of two two-level
QDs separated by a distance $d$, which are coupled to two
one-dimensional metal nanowires. The QD-1(2) is characterized by
the ground state $|g_{1(2)}\rangle$ and excited state
$|e_{1(2)}\rangle$. The two nanowires are connected by the two
distant QDs with the coupling strength $V_{nm}$($n=1,2;m=a,b$
throughout the paper) between nanowire SPs and QDs. For the
dispersion relations of nanowire SPs, we focus on the nonlinear
case, which exhibits a parabola-like curve with a local minimum for
the $1$th excitation mode~\cite{Chen8,Chen9}. In the parabola-like
frequency diagram, the dispersion relation $\omega_k$ around the
local minimum can be approximated as a quadratic form as follows:
\begin{equation}
\omega_k=E_k/\hbar=\omega_c+A(k-k_{0})^{2},
\label{eq:k}
\end{equation}
where $k$ is the wavevector of SP and $k_0$ is the wavevector for
the local minimum frequency $\omega_c$. Adopting the units in
Ref~\cite{Chang2}, the value of $A$ is about $0.0001$, and the
value of the reduced wavevector $K_0(= k_0c/\omega_p)$ around the
local minimum is about 15, where $\hbar\omega_p = 3.76 eV$ is the
plasma energy of the silver nanowire. For simplicity, we assume that
the two nanowires are identical and have the same frequency
$\omega_c$ and the same value of $A$. When a single SP is incident
from the left of the nanowire-a, it will propagate or be reflected
by the QDs along the four ports of the two nanowire channels, as
the system is served as a quantum router.

The real-space Hamiltonian~\cite{Chen5,Shen} describing the system is expressed as
follows:
\begin{eqnarray}
 && H/\hbar=\sum_{m=a,b}\int dx
{\bigg{\{} }(\omega_{c}+Ak_0^{2}) [C_{Rm}^{\dag}(x)C_{Rm}(x)
\nonumber\\
&&+C_{Lm}^{\dag}(x)C_{Lm}(x)]\nonumber\\
&& - 2Ak_0[-iC_{Rm}^{\dag}(x)\frac{\partial}{\partial
x}C_{Rm}(x)+iC_{Lm}^{\dag}(x)\frac{\partial}{\partial x}C_{Lm}(x)]\nonumber\\
&&+  A[\frac{\partial}{\partial
x}C_{Rm}^{\dag}(x)\frac{\partial}{\partial
x}C_{Rm}(x)+\frac{\partial}{\partial
x}C_{Lm}^{\dag}(x)\frac{\partial}{\partial x}C_{Lm}(x)]\nonumber\\
&&+ V_{1m}
 \delta(x)[C_{Rm}^{\dag}(x)|g_{1}\rangle\langle e_{1}|+C_{Lm}^{\dag}(x)|g_{1}\rangle\langle e_{1}|+H.c.]\nonumber\\
&&+  V_{2m}
 \delta(x-d)[C_{Rm}^{\dag}(x)|g_{2}\rangle\langle e_{2}|+C_{Lm}^{\dag}(x)|g_{2}\rangle\langle e_{2}|
+H.c.]{\bigg{\}}}
\nonumber\\&& + \sum_{n=1,2}(\omega_{n}-i\gamma_{n})|e_{n}\rangle\langle
e_{n}|,\label{eq:ham}
 \end{eqnarray}
where we have taken the energy level of the ground state
$|g_{n}\rangle$ as the energy reference. $C_R^{(\dag)}(x)$
$[C_L^{(\dag)}(x)]$ denotes the creation operator of a
right(left)-traveling surface plasmon in the nanowire at position
$x$. $\delta(x)[\delta(x-d)]$ indicates that the interaction occurs
at $x=0(d)$. $V_{nm}$ describes the coupling strength between SPs
and QDs. $\omega_n$ denotes the transition frequency of excited
state of the QD-n, and $\gamma_n$ the dissipation rate of the QD-n.
$|e_{n}\rangle\langle e_{n}|$ represents the diagonal element and
$|g_{n}\rangle\langle e_{n}|$ represents the off-diagonal of the nth
QD operator.

\begin{figure}[htbp]
\centering
\includegraphics[width=8cm,height=5cm]{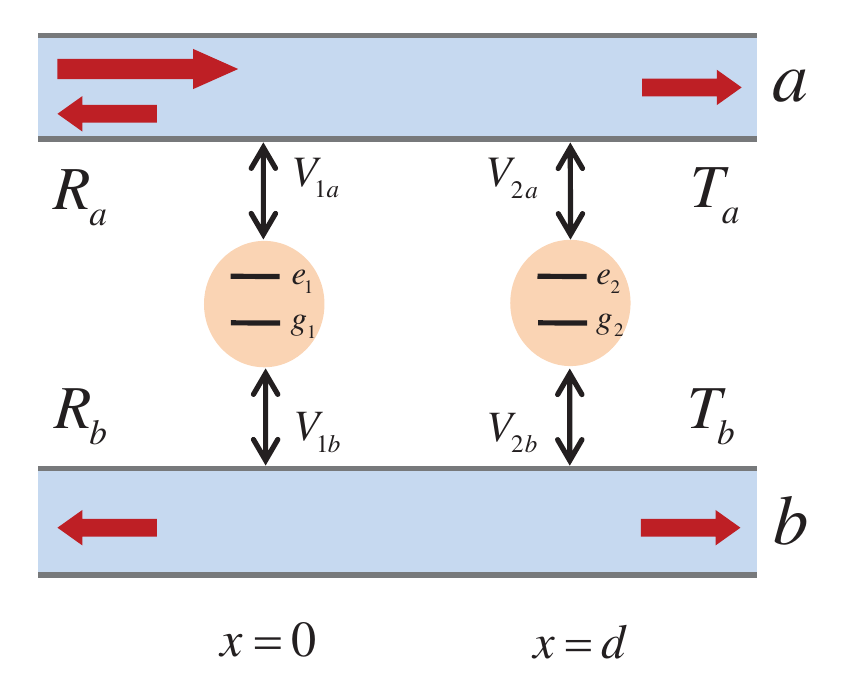}
\caption{(Color online) Schematic diagram for the quantum routing of
single SPs in two channels made of two nanowires coupled to two QDs.
Two QDs characterized by $|g_{n}\rangle$ and $|e_{n}\rangle (n=1,2)$
are located at $x=0$ and $x=d$, respectively. An incoming wave from
the left side of the nanowire-a will be reflected, transmitted, or
transferred into the nanowire-b.}\label{Fig.1}
\end{figure}

Assume that the QDs are initially in its ground states and a single
SP is incident from the left with the energy $E_k$. The scattering
eigenstate of the Hamiltonian (\ref{eq:ham}) is then given by
\begin{eqnarray}
|\psi\rangle &=& \sum_{m=a,b}\int  dx[ \phi_{Rm}(x) C_{Rm}^\dag(x) +
\phi_{Lm}(x) C_{Lm}^\dag(x)]
\nonumber\\&+&|\emptyset_{a},\emptyset_{b},g_{1},g_{2}\rangle+\xi_1|\emptyset_{a},\emptyset_{b},e_{1},g_{2}\rangle+
\xi_2 |\emptyset_{a},\emptyset_{b},g_{1},e_{2}\rangle \nonumber\\
\end{eqnarray}
where $\phi_{Rm(Lm)}$ is the single SP wave function in the
right(left) mode of the nanowire-m. $|\emptyset_{a},\emptyset_{b},
g_{1},g_{2}\rangle$ denotes that both QD-1 and QD-2 are in their
ground states with no SPs. $\xi_n(n=1, 2)$ is the probability
amplitude that the nth QD absorbs the surface plasmon and jumps to
its excited state.

The corresponding wave function can also be expressed as
\begin{eqnarray}
&&\phi_{Ra}(x)=e^{ikx}[\theta(-x)+t_{12}^{a}\theta(x)\theta(d-x)+t_{a}\theta(x-d)],\nonumber\\
&&\phi_{La}
(x)=e^{-ikx}[r_{a}\theta(-x)+r_{12}^{a}\theta(x)\theta(d-x)],\nonumber\\
&&\phi_{Rb}(x)=e^{ikx}[t_{12}^{b}\theta(x)\theta(d-x)+t_{b}\theta(x-d)],\nonumber\\
&&\phi_{Lb}(x)=e^{-ikx}[r_{b}\theta(-x)+r_{12}^{b}\theta(x)\theta(d-x)].
\end{eqnarray}
where $\theta(x)$ is the Heaviside step function with
$\theta(0)=1/2$. $t_{a(b)}$ and $r_{a(b)}$ are the transmission and
reflection amplitude in the metal nanowire-a(b), respectively.
$t_{12}^{a(b)}\theta(x)\theta(d-x)$ and
$r_{12}^{a(b)}\theta(x)\theta(d-x)$ represent the wave function of
the SP between QD-1 and QD-2.

Solving the eigenvalue equation $H|\psi\rangle=E_k|\psi\rangle$, one
get the exact forms of the reflection and transmission coefficients
as follows:
\begin{widetext}
\begin{eqnarray}
 &&t_a=\frac{B(k)(V_{1b}^{2}E_{k2}+V_{2b}^{2}E_{k1})+B(k)^2E_{k1}E_{k2}
 +V_{1b}V_{2b}[1-e^{i\phi(k)}]M}{B(k)(Q_{1}E_{k2}+Q_{2}E_{k1})+B(k)^2E_{k1}E_{k2}
 +[1-e^{i\phi(k)}]M^{2}+N^{2}},\nonumber\\
 &&r_a=\frac{-B(k)[V_{1a}^{2}E_{k2}+V_{2a}^{2}e^{i\phi(k)}E_{k1}]-[1-e^{i\phi(k)}]V_{1a}^{2}V_{2a}^{2}
 +V_{2a}V_{1b}e^{i\phi(k)}(V_{1a}V_{2b}+N)-V_{1a}^{2}V_{2b}^{2}}{B(k)(Q_{1}E_{k2}+Q_{2}E_{k1})
 +B(k)^2E_{k1}E_{k2}+[1-e^{i\phi(k)}]M^{2}+N^{2}},\nonumber\\
 &&t_b=\frac{-B(k)(V_{1a}V_{1b}E_{k2}+V_{2a}V_{2b}E_{k1})-[1-e^{i\phi(k)}]V_{2a}V_{1b}M}{B(k)(Q_{1}E_{k2}
 +Q_{2}E_{k1})+B(k)^2E_{k1}E_{k2}+[1-e^{i\phi(k)}]M^{2}+N^{2}},\nonumber\\
 &&r_b=\frac{-B(k)[V_{1a}V_{1b}E_{k2}+V_{2a}V_{2b}e^{i\phi(k)}E_{k1}]
 -[1-e^{i\phi(k)}]V_{1a}V_{1b}Q_{2}}{B(k)(Q_{1}E_{k2}+Q_{2}E_{k1})+B(k)^2E_{k1}E_{k2}+[1-e^{i\phi(k)}]M^{2}+N^{2}}
 \label{eq:tr}
\end{eqnarray}
\end{widetext}
 where $Q_{1}=V_{1a}^{2}+V_{1b}^{2}$,
$Q_{2}=V_{2a}^{2}+V_{2b}^{2}$, $M=V_{1a}V_{2a}+V_{1b}V_{2b}$,
$N=V_{1a}V_{2b}-V_{1b}V_{2a}$,
$E_{kn}=A(k-k_{0})^{2}+\delta_{n}-i\gamma_{n}$,
$\delta_{n}=\omega_c-\omega_n$, $\phi(k)=2kd$, and
$B(k)=iA(2k_0-k)$.
The transmission amplitude $T_{a(b)}=|t_{a(b)}|^2$ and reflection
amplitude $R_{a(b)}=|r_{a(b)}|^2$ can then be determined
algebraically.

\section{Quantum routing of the SPs}

The routing property of single SPs is characterized by the
transmission and reflection coefficients of these forts. It can be checked numerically for the
relation $T_a+R_a+T_b+R_b=1$ in the nondissipative case
$(\gamma_1=\gamma_2=0)$, which means that the transmission and
reflection probabilities are conserved.

As a contrast, we first discuss the scattering properties of SPs in
the single-dot case, in which the inter-dot distance $d$ in Fig.~1
is taken to be in the limit of $d\rightarrow 0$. As shown in Fig.~2, the transmission in four ports of the incident SP is affected slightly when the detuning ($\delta=\delta_1=\delta_2$) is positive, while two peaks (dips) can emerge for negative detuning. To explain this clearly, we reduce the
Eq.~(\ref{eq:tr}) and get the reduced forms of $t_a=[2V^2+B(k)E_{k1}]/[4V^2+B(k)E_{k1}]$ and
$r_a=t_b=r_b=2V^2/[4V^2+B(k)E_{k1}]$, with the same coupling $V$.
Obviously, for negative detuning in $E_{k1}$ under the quadratic dispersion relation, it yields two different values corresponding to the two dips (peaks) in the
curves.

\begin{figure}[htbp]
\centering
\includegraphics[width=8cm,height=6cm]{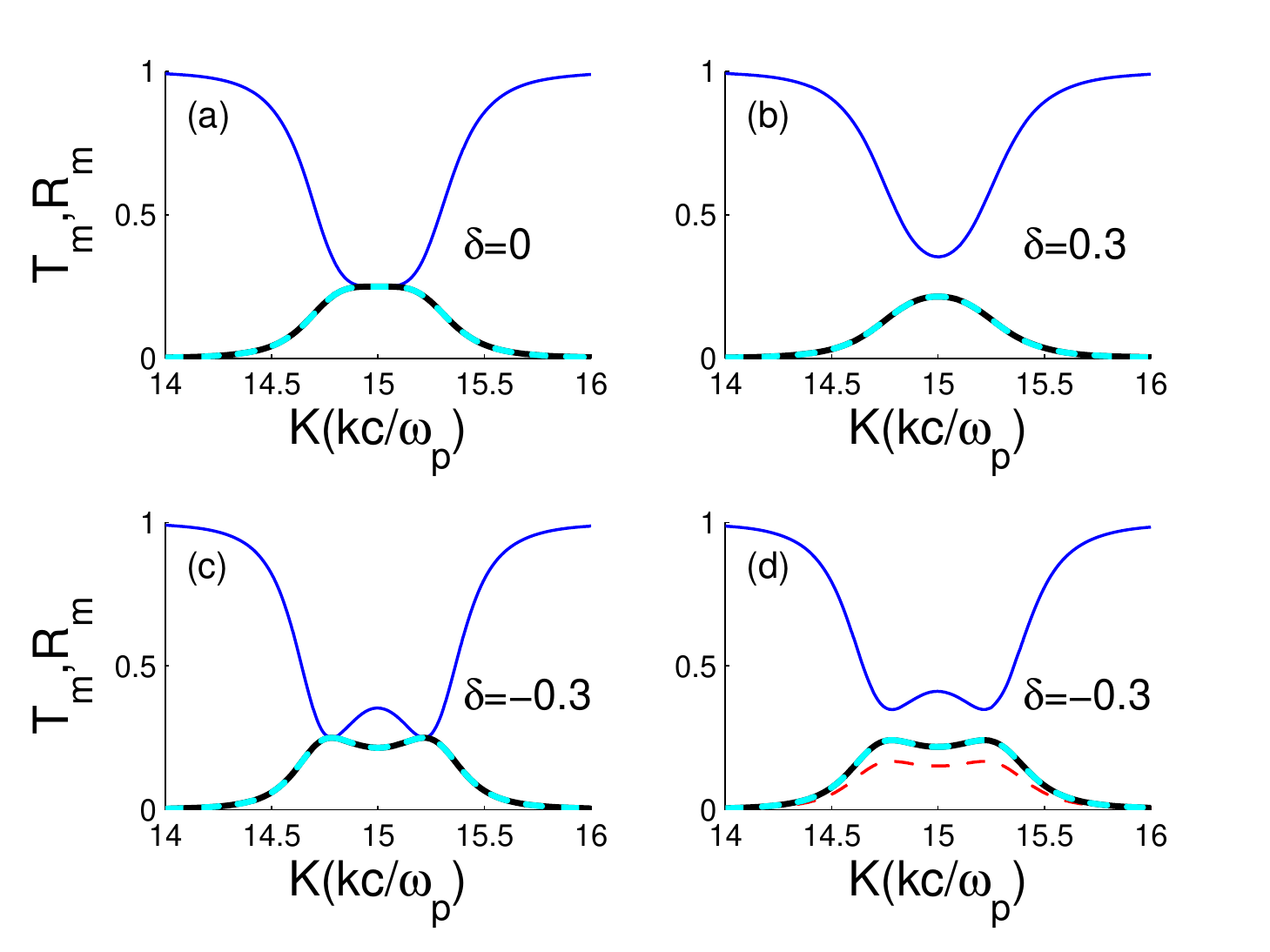}
\caption{(Color online) Transmission $T_m$ (thin solid blue $T_a$
and thick solid black $T_b$ ) and reflection $R_m$ (thin dashed red
$R_a$ and thick dashed cyan $R_b$) spectra of the single-QD case for
different detunings $\delta$ (a) $\delta=0$, (b) $\delta=0.3$, (c)
$\delta= -0.3$, and (d) $\delta= -0.3$, with $V_a =V_b =1$ in (a),
(b), and (c), and $V_b = 1.2V_a $ in (d). Here, the detuning is
defined as: $\delta=\delta_1=\delta_2$. In (a), (b), and (c), the
curves of $R_a$, $T_b$, and $R_b$ overlap each other, while $R_a$
separates in (d) for different couplings. Throughout the paper,
$\delta$ is in unit of $10^{-6}\omega_p$, $V_a$ and $V_b$ are in the
same unit of $1.5\sqrt{10^{-10}\omega_pc/4\pi}$, and $k_0$ is in unit of
$\omega_p/c$. }\label{Fig.2}
\end{figure}

To describe the transfer rate of Sps from the input channel into another channel, the parameters $P_a =T_a+R_a$ and $P_b =T_b+R_b$ are introduced to
present the probabilities of finding the incident single SPs in
nanowire-a and nanowire-b, respectively. Fig.~3 exhibits the $P_a$
and $P_b$ for different inter-dot distance $d$ with $V_a =V_b= 1$
and $\delta_1=\delta_2=0$, respectively. $P_b\leq 0.5$ indicates
that a single SP incident from the input channel can be redirected
into another channel with a maximum probability of 0.5, which
is similar to the single-emitter case investigated in
Refs.~\cite{Zhou,Lu}. Interestingly, when the two QDs are separated, the transfer rate
$P_b$ can be more than 0.5 in a wide rang of wavenumbers near $k_0$.
Indeed, as seen in Fig.~3 (d), a wide region of $P_b>0.5$ is
exhibited when $d>0$, and the transfer rate $P_b$ is a periodic
function of the distance $d$ related to the phase shift $\phi$. This
indicates that the present double-dot configuration may provide an
effective approach to enhance the routing capability of the SPs.
\begin{figure}[htbp]
\centering
\includegraphics[width=8cm,height=6cm]{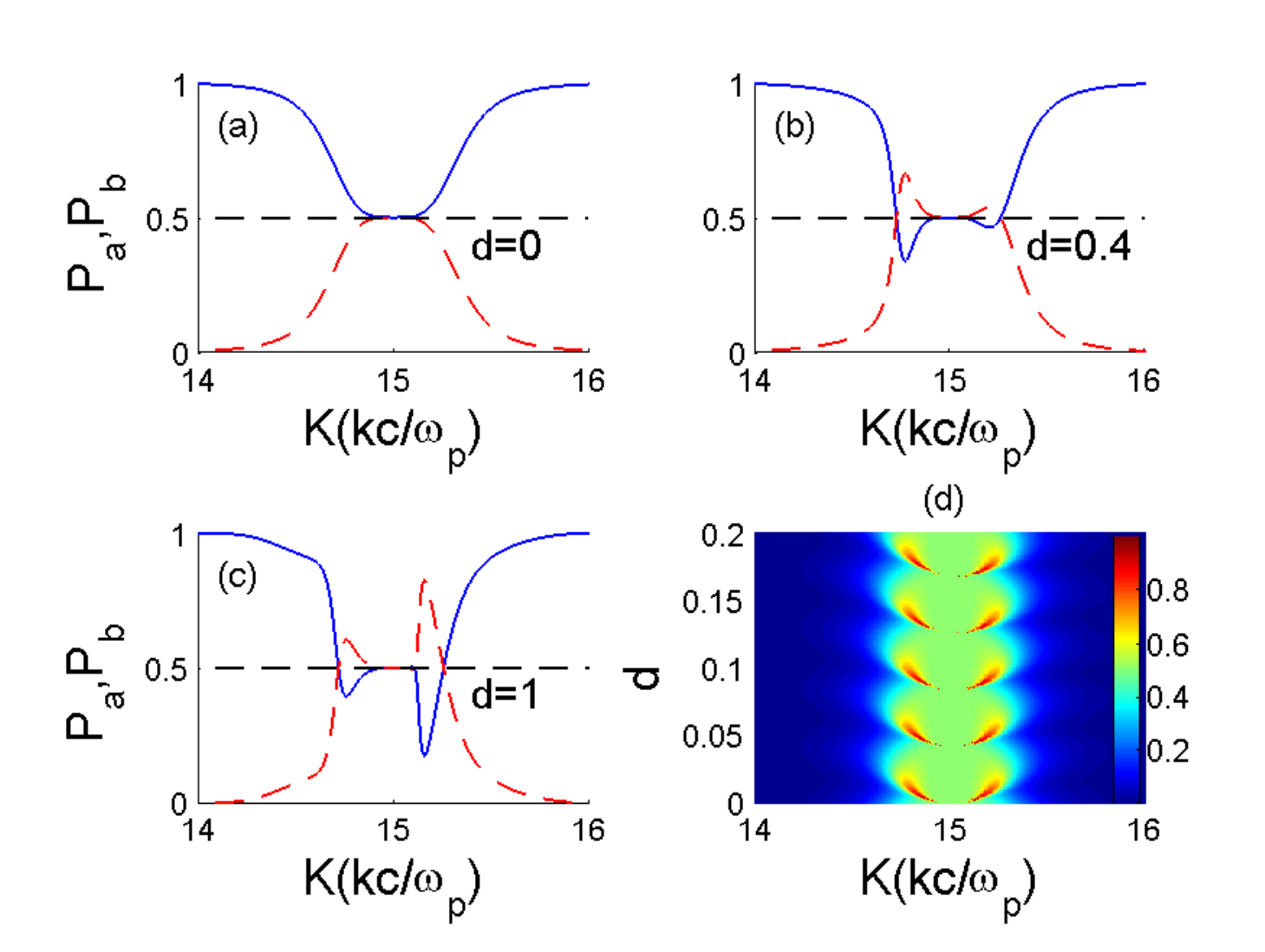}
\caption{(Color online) $P_a$ (solid blue) and $P_b$ (dashed red)
for different inter-dot distance (a) $d=0 $, (b) $d=0.4 $, (c) $d=1
$, respectively. (d) $P_b$ as functions of the distance $d$ and the
reduced wavenumber $K$. Other parameters are $\delta_1=\delta_2=0$,
and $V_a =V_b= 1$. Throughout the paper, $d$ is in unit of
$5c/\omega_p$.}\label{Fig.3}
\end{figure}
\begin{figure}[htbp]
\centering
\includegraphics[width=8cm,height=6cm]{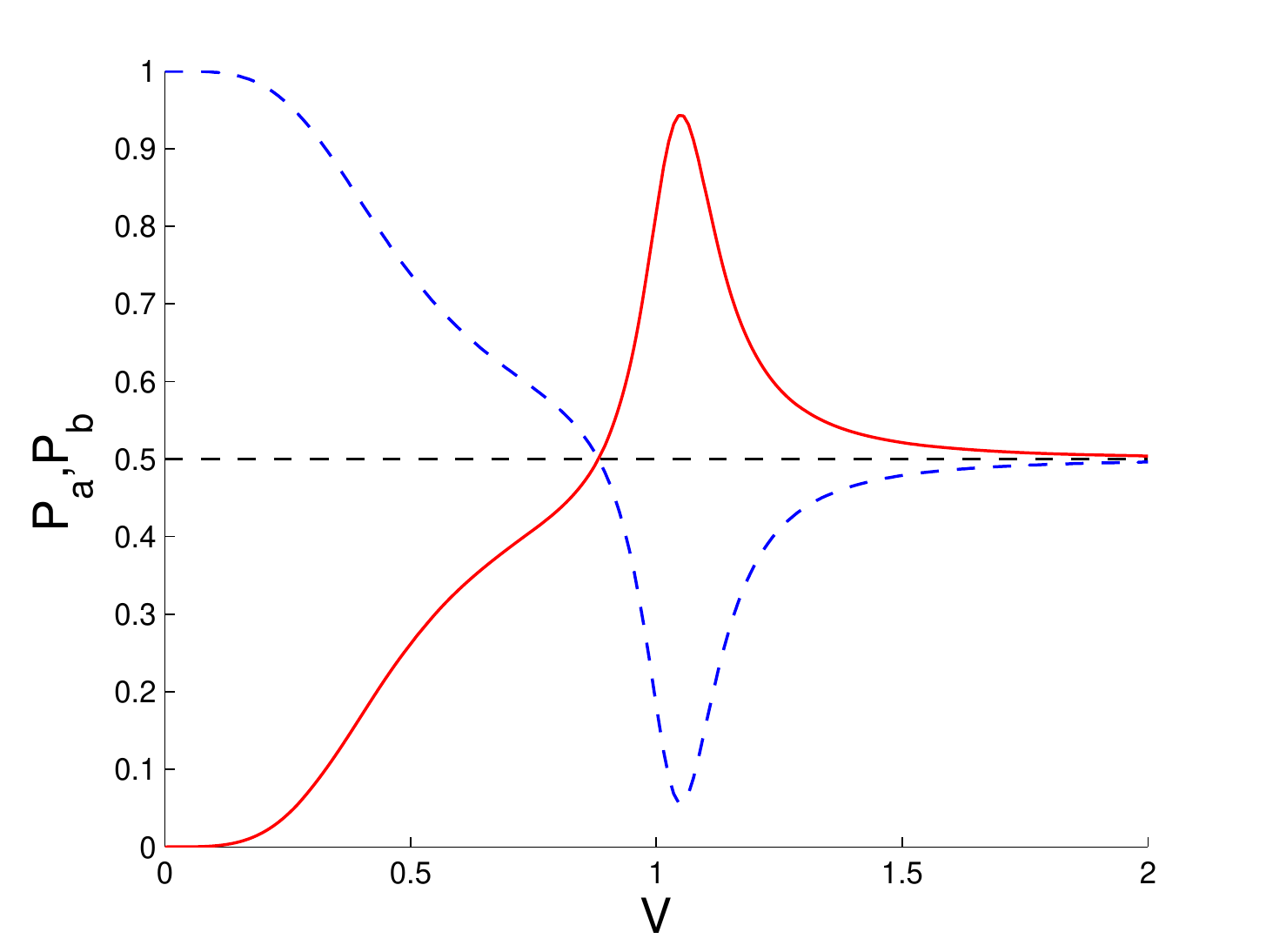}
\caption{(Color online) The influence of QD-SP coupling on the $P_a$
(dashed blue) and $P_b$ (solid red).  These parameters are
$\delta_1=\delta_2=0$, $d=1$, and $k=1.01k_0$.  }\label{Fig.4}
\end{figure}

Furthermore, to investigate the dependence of the transfer rate
of the single SPs on the QD-SP coupling, in Fig.~4 we plot $P_b$ and
$P_a$ versus the coupling $V=V_a=V_b$ at a fixed wavevector
$k=1.01k_0$. As seen, when the coupling is tuned off, the
input SPs is almost completely transmitted, with the transmission $T
=1$ in the right-going direction. While the coupling is
switched on, the transfer rate from nanowire-a to nanowire-b
can reach the maximum value (i.e., $P_b\approx0.95$), and thus high
transfer rate is realized. The transfer rate $P_b$ is also
plotted for the different couplings in Fig.~5, which describes the
quantum routing in detail. One can see that, when the couplings
$V_a$ and $V_b$ are sufficiently small, the input SPs cannot be
redirected into another channel completely. When increasing the couplings, a large
region of high transfer appears, and a perfect transfer rate
($P_b\approx 1$) is achieved when the $V_a=V_b$ is designed as
around $1.05$. Therefore, it is possible to efficiently control the
routing capability of the single SPs from the a channel into another
channel via designing the QD-SP coupling accordingly.

\begin{figure}[htbp]
\centering
\includegraphics[width=8cm,height=6cm]{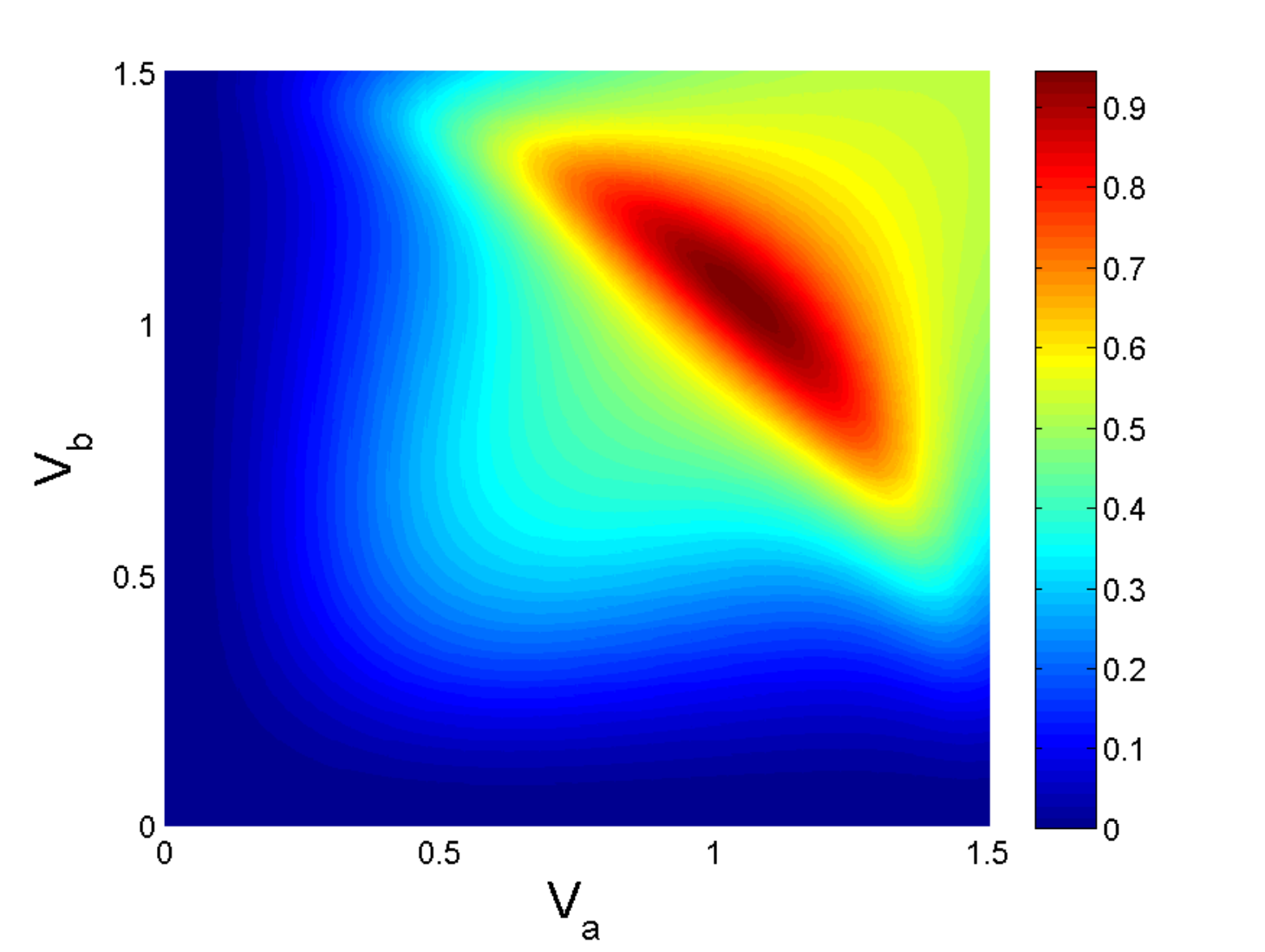}
\caption{(Color online) The influence of different couplings on the
$P_b$.  These parameters are $\delta_1=\delta_2=0$, $d=1$, and
$k=1.01k_0$. }\label{Fig.5}
\end{figure}

To gain a deeper insight into the dependence of the routing
capability of SPs on more controllable parameters, Fig.~6 displays
the transfer rate $P_b$ of the single SPs versus the QD-SP
coupling and the phase shift $\phi$ related to the distance $d$.
We can find that the transfer rate $P_b$
is a periodic function of the phase shift $\phi$, and
there is a wide window of $P_b>0.5$ when increasing the QD-SP
coupling. For high transfer rate from one channel to the other,
a long and narrow region emerges.

\begin{figure}[htbp]
\centering
\includegraphics[width=8cm,height=6cm]{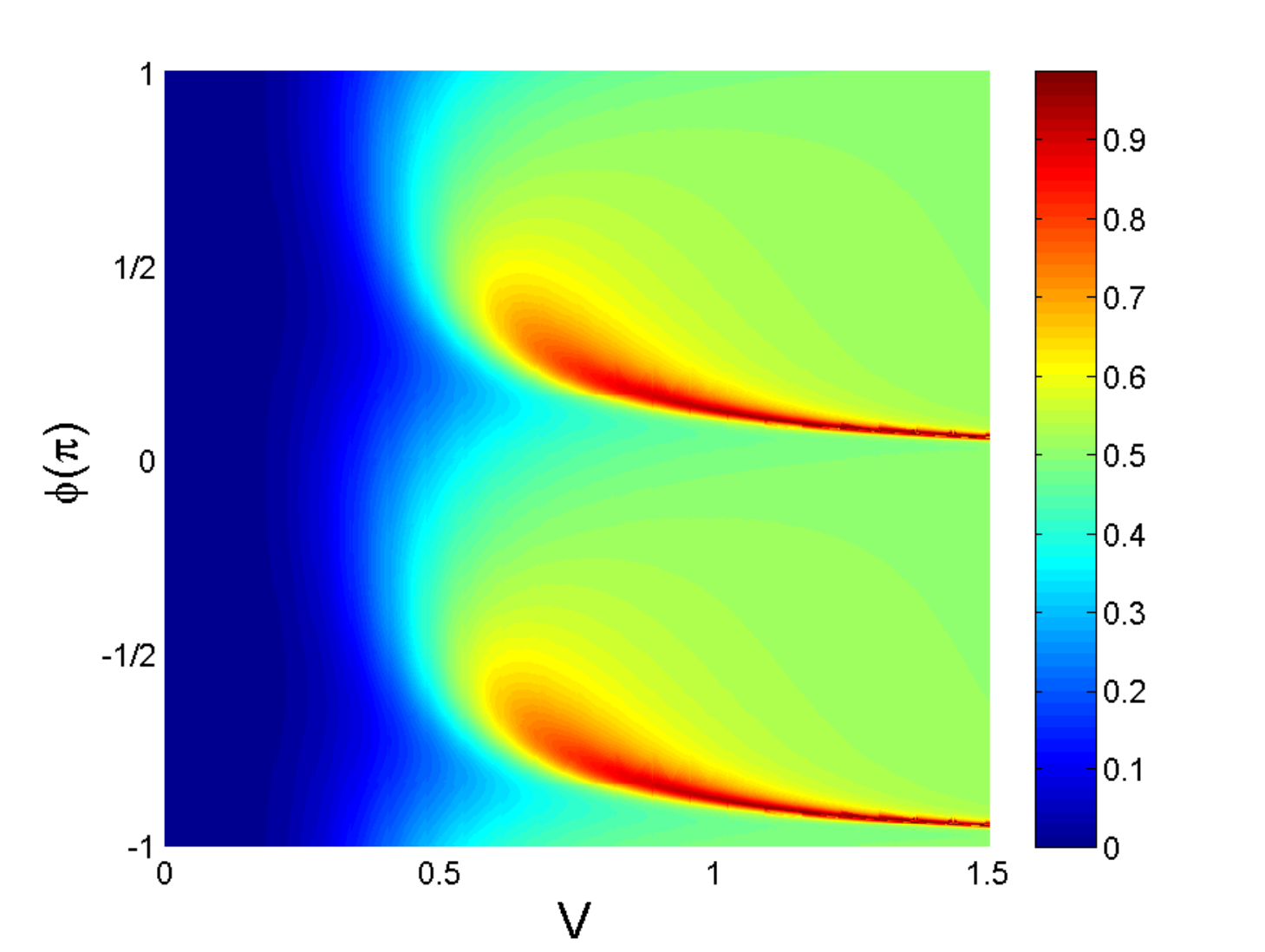}
\caption{(Color online) The influence of coupling and phase shift on
the $P_b$.  The relevant parameters are set as:
$\delta_1=\delta_2=0$, $V_a =V_b=V$, and $k=1.01k_0$. }\label{Fig.6}
\end{figure}

\section{Fano-like resonances of the SPS}

With the further increase of $d$, the jiggling behavior in the scattering spectra emerges and becomes more obvious, as shown in Fig.~7(c) and (d). The oscillations may
originate from the multiple interference of waves in the region
between the two QDs~\cite{Zhou2}. As the inter-resonator distance
increases, more wave interference results in much larger oscillations of the scattering spectra.

\begin{figure}[htbp]
\centering
\includegraphics[width=8cm,height=6cm]{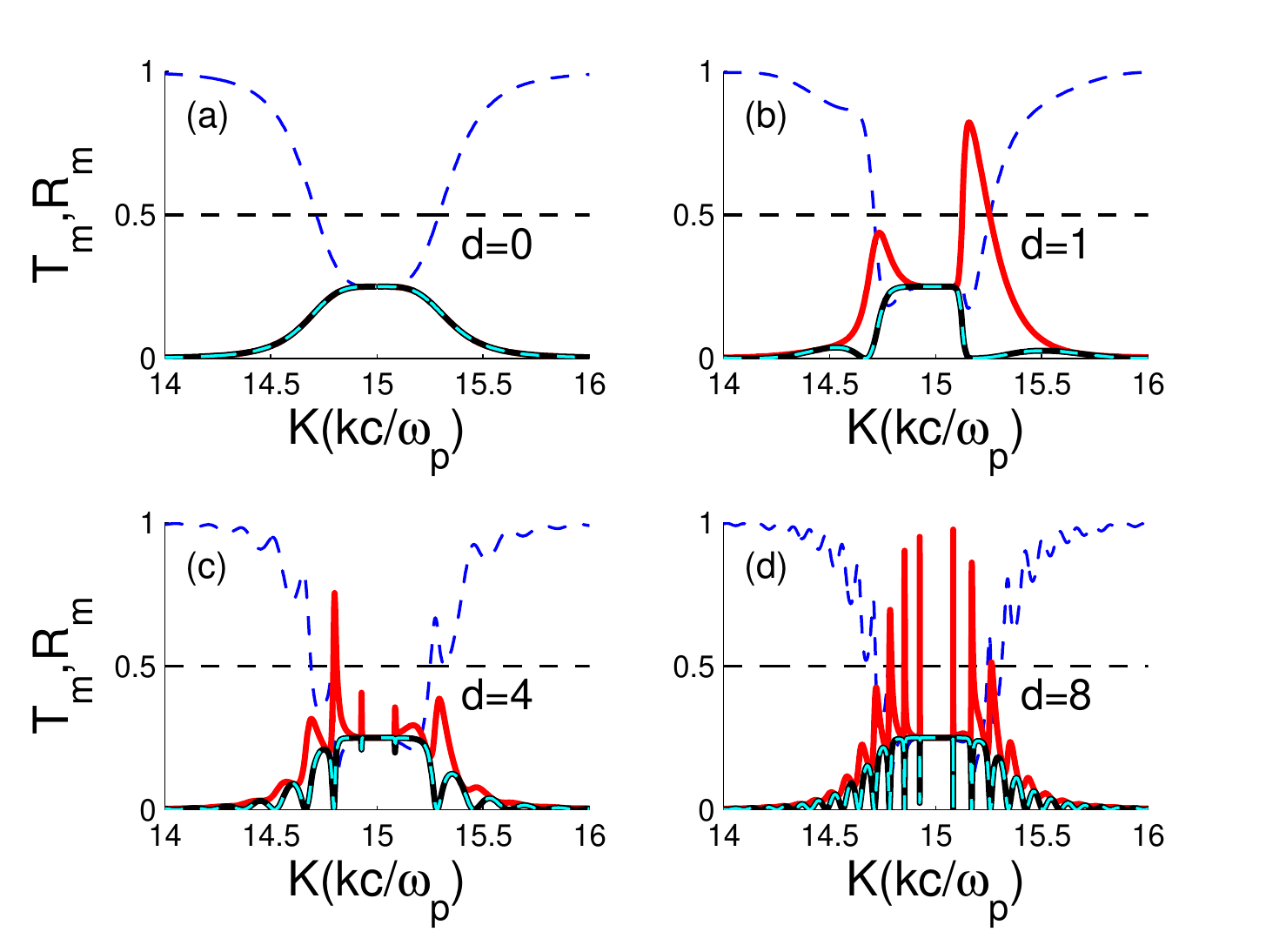}
\caption{(Color online) Fano-like resonances of transmission ( thin
dashed blue $T_a$ and  thick solid red $T_b$ ) and reflection (
thick solid black $R_a$ and thin dashed cyan $R_b$) spectra for
different inter-dot distance in (a) $d =0$, (b) $d =1$, (c) $d =4$, and (d) $d =8$, respectively. Other parameters
are $\delta_1=\delta_2=0$, and $V_a =V_b= 1$. }\label{Fig.7}
\end{figure}

More interestingly, the feature of Fano resonance in the jiggling
spectra becomes also more evident when increasing $d$. Obviously,
Fano-line shapes don't emerge for the single-dot case ($d=0$), whereas
Fano resonance occurs at $d=4$. In particular, more sharp peaks
(dips) of Fano-line shapes are generated by the nonlinear dispersion
relation at $d=8$ for the the transmission (reflection) spectra.

Specifically, to mathematically illustrate the feature related to Fano resonance, we
consider one of the reflection coefficients as an example.
When $V_a=V_b=V$ and $\delta_1=\delta_2=\delta$, the
reduced form of the reflection coefficient $R_b$ for the quadratic
dispersion relation (\ref{eq:k}) is expressed as
\begin{eqnarray}
R_b={\bigg|}\frac{B(k)V^2E_{k1}[1+e^{i\phi(k)}]
 +2V^4[1-e^{i\phi(k)}]}{4B(k)V^2E_{k1}+B(k)^2E_{k1}^2+2V^{2}[1-e^{i\phi(k)}]}{\bigg|}^2.
 \label{eq:f1}
\end{eqnarray}
Obviously, the zero value of $R_b$ emerges when two functions
\begin{eqnarray}
X= A(2k_0-k)[A(k-k_{0})^{2}+\delta]\cos{(kd)}
\label{eq:f2}
\end{eqnarray}
and
\begin{eqnarray}
Y= 2V^2\sin{(kd)}
\end{eqnarray}
coincide with each other. Fig.~8(a) displays multiple Fano-line shapes for the reflection spectra of $R_b$ under the parameter conditions: $d = 8$, $V=1$, and $\delta=
0$. Fig.~8(b) demonstrates that many intersections of functions $X$
(thin blue curve) and $Y$ (thick red solid curve) represent
correspondingly the zeros of $R_b$, where multiple Fano-line shapes
appear in the jiggling spectra. From Eqs.~(\ref{eq:f1}) and (\ref{eq:f2}), one knows that a number of intersections of functions occur under the nonlinear condition, which reveals the appearance of multiple Fano-like resonance in the case is due to the quadratic dispersion relation.

Fano-like resonances in such similar manners have been investigated
in recent works~\cite{Chen5,Chen6,Chen7}. Physically, in the present double-dot
configuration, QD-1 (QD-2) is served as a delocalized (localized)
channel for the SPs passing through it. Consequently, the
interference between the localized and delocalized channels of the
incident SPs results in asymmetric line shapes around these
double-peak profiles. Increasing the
inter-dot distance would enhance the interference between the two
channels and thus lead to more distinct Fano-line shapes.

\begin{figure}[htbp]
\centering
\includegraphics[width=8cm,height=6cm]{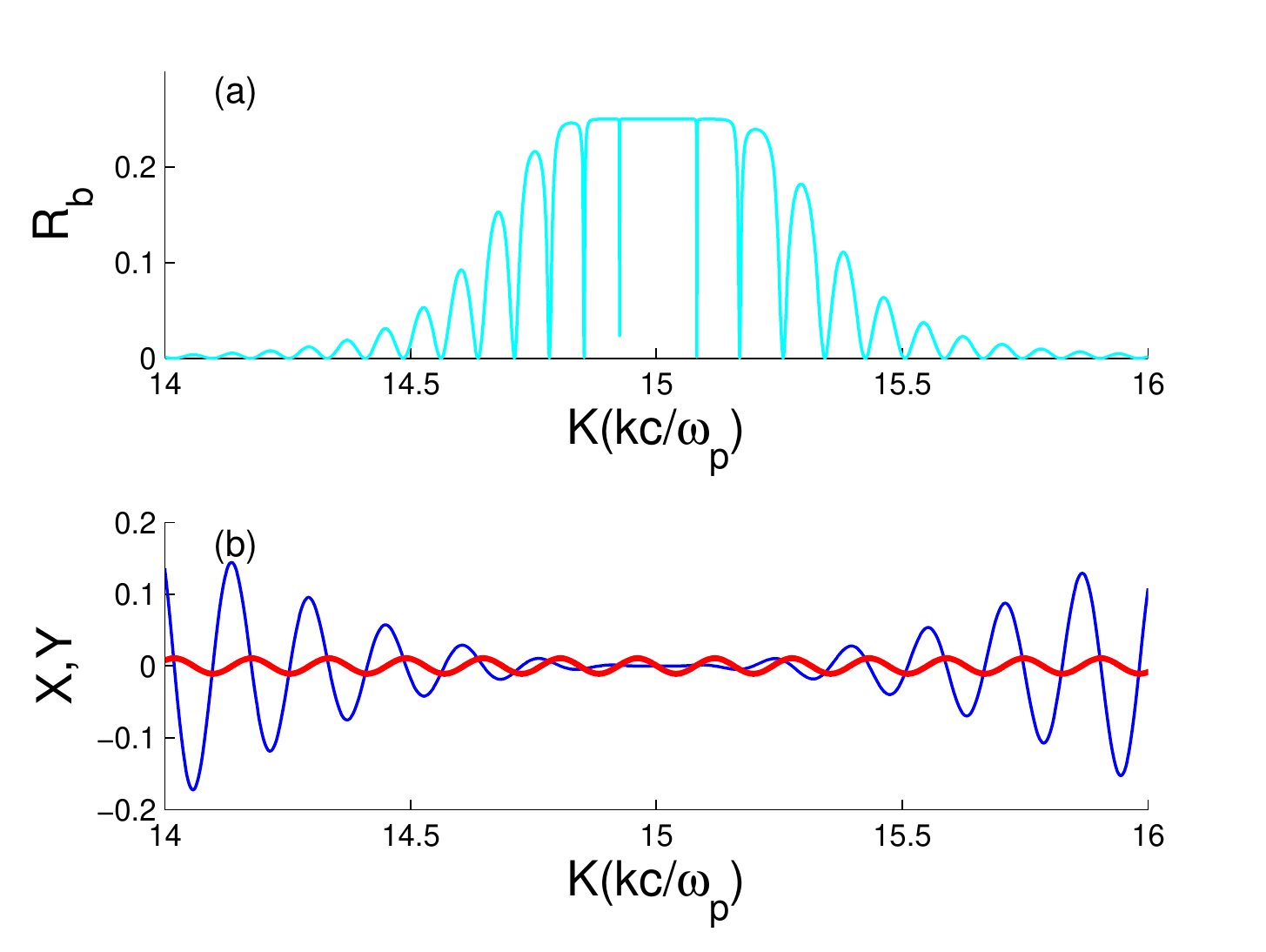}
\caption{(Color online) (a) Reflection spectra $R_b$ with $d = 8$,
$V_a=V_b=V=1$, and $\delta_1=\delta_2=\delta=0$. (b) The
intersections of functions $X$ (thin blue curve) and $Y$ (thick red
curve) represent the zeros of $R_b$. Here  $X$ and $Y$ are in the
same unit of $\omega_p$. }\label{Fig.8}
\end{figure}

\section{Conclusion }

In summary, by using the real-space Hamiltonian with nonlinear dispersion relation, we investigate the routing properties of the SPs propagating on the surface of two
metal nanowires coupled to two QDs. We find that, for the single-dot case, the transport of the incident SPs exhibit two peaks and dips in the scattering spectra for negative detunings, due to the quadratic dispersion of the waveguides. While for the double-dot case, the jiggling behavior and distinct Fano-like resonance emerge in the scattering spectra. More importantly, our results show that the routing capability of the SPs inputting from one channel to another can be enhanced to exceed 0.5. Sufficiently high transfer rate is also available by properly designing the inter-dot distance and the QD-SP coupling.

Noted that the dissipations are not considered in our
investigations. In real experiments, the QD-SP system inevitably
experience dissipations, such as the excited-state relaxation of the quantum dots
and plasmon loss, since it is also coupled to the environment. However, in
contrast with the strong QD-SP coupling, the dissipative scale of
the QD exciton ($\hbar$/lifetime $\approx 1\mu eV$) is smaller by two to three
orders of magnitude. Therefore, the influence of the
dissipations to the routing properties demonstrated above should
not be dominant, although it may lower a little the probabilities of
the scattering coefficients.

\acknowledgments
This work was supported by the National Natural Science Foundation
of China (Grant Nos. 11247032 and U1330201) and by the Natural
Science Foundation of Jiangxi (Grant No. $20151$BAB$202012$).


\end{document}